\documentclass[10pt,conference]{IEEEtran}

\usepackage{amsfonts}
\usepackage{setspace}

\usepackage{setspace}
\usepackage{amssymb}
\usepackage{epsfig}
\usepackage{graphicx}
\usepackage{xcolor}
\usepackage{algorithm}
\usepackage{algorithmicx}
\usepackage{algpseudocode}
\usepackage{amsmath}

\floatname{algorithm}{Algorithm}

\usepackage[ps2pdf,
bookmarks=false,
bookmarksnumbered=false, 
bookmarksopen=false, 
colorlinks=false]{}

\begin{document}

\title{A Deep Reinforcement Learning-Based Framework for Content Caching}

\author{Chen Zhong, M. Cenk Gursoy, and Senem Velipasalar
\\Department of Electrical Engineering and Computer Science,
Syracuse University, Syracuse, NY 13244
\\Email: czhong03@syr.edu, mcgursoy@syr.edu, svelipas@syr.edu}

\maketitle

\begin{abstract}
Content caching at the edge nodes is a promising technique to reduce the data traffic in next-generation wireless networks. Inspired by the success of Deep Reinforcement Learning (DRL) in solving complicated control problems, this work presents a DRL-based framework with Wolpertinger architecture for content caching at the base station. The proposed framework is aimed at maximizing the long-term cache hit rate, and it requires no knowledge of the content popularity distribution. To evaluate the proposed framework, we compare the performance with other caching algorithms, including Least Recently Used (LRU), Least Frequently Used (LFU), and First-In First-Out (FIFO) caching strategies. Meanwhile, since the Wolpertinger architecture can effectively limit the action space size, we also compare the performance with Deep Q-Network to identify the impact of dropping a portion of the actions. Our results show that the proposed framework can achieve improved short-term cache hit rate and improved and stable long-term cache hit rate in comparison with LRU, LFU, and FIFO schemes. Additionally, the performance is shown to be competitive in comparison to Deep Q-learning, while the proposed framework can provide significant savings in runtime.
\end{abstract}


\section{Introduction}

Recent years have witnessed rapid developments in rich media-enabled applications on mobile devices, providing users easier access to high-quality multimedia contents. With this, users increasingly demand more multimedia content, which usually has larger data size and requires more resources for transmission. As a consequence, the contents that need to be streamed in real-time have grown rapidly in terms of size and volume, which has led to the congested data traffic at content servers and degradation in user experience.

Content caching is a promising technique that enables data offloading and alleviates the data traffic congestion. In particular, this technique is aimed at pre-caching the contents at the end users or the base stations from the content servers. In this way, the time and resources needed to request and transport contents from upper level content servers or original content servers can be effectively saved, and data traffic can be reduced.

However, with content caching arises the policy control problem, in which we have to explore and decide which contents to store in caches.
Inspired by the success of Deep Reinforcement Learning (DRL) in solving complicated control problems, we in this paper design a DRL agent for content caching decisions at an edge node, e.g., a base station.
End-users keep requesting content from the base station and these requests are the input of the system. We suppose there is a queue of requests to be served.
The base station has a fixed storage capacity $C$, which is the maximum number of contents that can be cached at this base station.

The main contributions of this paper are summarized below:
\begin{itemize}
	\item For the first time, we present a deep reinforcement learning framework (that utilizes Wolpertinger architecture) for content caching problem. We define the state and action spaces and the reward function for the DRL agent, and employ this agent to make proper cache replacement decisions to maximize the cache hit rate.
	\item We analyze the performance of this DRL agent in terms of the cache hit rate. And we compare the performance with other caching algorithms, including Least Recently Used (LRU), Least Frequently Used (LFU), and First-In First-Out (FIFO) caching strategies. The results show that the DRL agent is able to achieve improved short-term cache hit rate and improved and stable long-term cache hit rate.
	\item We further confirm the effectiveness of the DRL agent through comparisons with deep Q-network. The results show that the DRL agent is able to achieve competitive cache hit rates while having significant advantages in runtime.
	
\end{itemize}

\section{Related Work}
Content caching has attracted much interest recently. For instance, the authors in \cite{kader2015leveraging} analyzed proactive content caching from the aspect of big data analytics and demonstrated that with limited cache size, proactive caching can provide $100\%$ user satisfaction while offloading $98\%$ of the backhaul traffic. In \cite{blasco2014learning}, three learning-based content replacement algorithms were studied with each leading to different exploitation-exploration trade-offs. The study in \cite{leconte2016placing} proposed an age-based threshold policy which caches all contents that have been requested more than a threshold. Furthermore, popularity-based content caching policies named StreamCache and PopCaching were studied in \cite{li2016streamcache} and \cite{li2016popularity}, respectively. More recently, increasing attention has been cast on machine learning based methods. In \cite{song2017learning}, content popularity is estimated using the multi-armed bandit algorithm. And the authors in \cite{tanzil2017adaptive} proposed an extreme-learning machine framework for content popularity prediction.

As seen in previous studies, content popularity distribution is always the key to solve the content caching problem. Considering the large scale of contents and the changing popularities, we note that deep reinforcement learning is a particularly attractive strategy to tackle this problem. In \cite{gruslys2017reactor}, \cite{lillicrap2015continuous} and \cite{dulac2015deep}, reinforcement learning methods are addressed. For instance, the proposed reinforcement learning agent in \cite{gruslys2017reactor} is a typical multi-step return actor-critic architecture. The deep deterministic policy gradient proposed in \cite{lillicrap2015continuous} describes how the network is to be updated. And the Wolpertinger policy proposed in \cite{dulac2015deep} provides a solution for large discrete action spaces.

\section{System Model}
Data traffic is triggered by requests from the rapidly increasing number of end-users, and the volume of requests varies over time. In this setting, we propose a deep reinforcement learning framework acting as an agent. Based on the users' requests, this DRL agent makes caching decisions to store the frequently requested contents at local storage. If the requested contents are already cached locally, then the base station can serve the user directly with reduced delay. Otherwise, the base station requests these contents from the original server and updates the local cache based on the caching policy.

In this work, we consider a single base station with cache size of $C$. We assume that in a given time slot, the total number of contents that users can request from this base station is fixed and denoted as $N$. We give every content a unique index, and this index acts as the content ID. We assume that all contents have the same size. The list of users' requests is denoted as $Req = \{R_{1}, R_{2}, R_{3},...\}$. Here, $R_{t}$ denotes the ID of the requested content at time $t$. For each request, the DRL agent makes a decision on whether or not to store the currently requested content in the cache, and if yes, the agent determines which local content will be replaced.

We define $\mathcal{A}$ as the action space, and let $\mathcal{A} = \{a_1,a_2,a_3,..., a_m\}$, where $a_{\upsilon}$ denotes a valid action. And in our case, $m$ has a finite but generally a large value, describing the total number of possible actions.  For each content, there are two cache states: cached, and not cached. The cache state gets updated based on the caching decision. Here, we define two types of actions: the first one is to find a pair of contents and exchange the cache states of the two contents; the second one is to keep the cache states of the contents unchanged. Theoretically, multiple actions can be executed at one decision epoch. To reduce the computational complexity, we need to limit the action space size $m$ and the number of actions to be executed in one decision epoch, which will discussed in detail in Section \ref{sec:framework}.

The reward should reflect the objective of the framework, which, in our case, is to reduce the data traffic. In our setting, all requests are served by the base station, all contents have the same size, and there are no priorities for users. Therefore, the reduction in data traffic can be evaluated in terms of the cache hit rate. Here, we define the cache hit rate $CHR$ in $T$ requests as
\begin{equation}
CHR_T = \frac{\sum_{i = 1}^{T}\mathbf{1}\left( R_i\right) }{T}
\end{equation}
where indicator function $\mathbf{1}\left( R_i\right)$ is defined as
\begin{eqnarray}\mathbf{1}\left( R_i\right)=
\begin{cases}
1, &R_i \in \mathcal{C}_T, \cr 0 & R_i \notin \mathcal{C}_T\end{cases}
\end{eqnarray}
where $\mathcal{C}_T$ stands for the cache state in this period.
Therefore the reward in $T$ requests can be defined as
\begin{equation}
r^T = CHR_T.
\end{equation}
For each decision epoch $t$, we obtain reward $r_t$, which can be a weighted sum of short-term and long-term cache hit rates. We more explicitly introduce the definition of $r_t$ for the proposed framework in Section \ref{sec:framework} below.

The objective of the DRL agent is to find a policy, $\sigma^*$, that maximizes the long-term cache hit rate:\
\begin{equation}\label{eq:eq1}
	\underset{\sigma^*}{\text{maximize}} \quad E[r_t|\sigma^*].
\end{equation}
\\
We are interested in developing a model-free learning algorithm to solve problem (\ref{eq:eq1}) that can effectively reduce the data traffic with fixed cache capacity at the base station.

\section{DRL-based Content Caching Framework}\label{sec:framework}

In this section, we present the DRL-based content caching framework, which is aimed at maximizing the cache hit rate in order to reduce the data traffic. To solve the content caching problem with high-dimensional state and action spaces (due to the large number of contents and cache sizes in practical scenarios), we propose a framework based on the Wolpertinger architecture \cite{dulac2015deep} to narrow the size of the action space and avoid missing the optimal policy at the same time.

\subsection{Algorithm Overview}\label{sub:algorithm}
Based on the Wolpertinger Policy \cite{dulac2015deep}, our framework consists of three main parts: actor network, K-nearest neighbors (KNN), and critic network. We train the policy using the Deep Deterministic Policy Gradient (DDPG) \cite{lillicrap2015continuous}. The Wolpertinger architecture is employed for two reasons: 1) as an online algorithm, this framework can adapt to data, and enables us to develop a long-term policy; 2) actor network can avoid the listing and consideration of very large action space, while the critic network can correct the decision made by the actor network, and KNN can help to expand the actions to avoid poor decisions. This algorithm work in three steps. Firstly, the actor network takes cache state and the current content request as its input, and provides a single proto actor $\hat{a}$ at its output. Then, KNN receives the single actor $\hat{a}$ as its input, and calculate the $l_2$ distance between every valid action and the proto actor in order to expand the proto actor to an action space, denoted by $\mathcal{A}_k$, with $K$ elements and each element being a possible action $a_{\upsilon} \in \mathcal{A}$. And at the last step, the critic network takes the action space $\mathcal{A}_k$ as its input, and refines the actor network on the basis of the $Q$ value. The DDPG is applied to update both critic and actor networks.

Below we provide a more detailed description of the key components of the algorithm.

\emph{The actor:}
We define the actor as a function parameterized by $\theta^{\mu}$, mapping the state $\mathcal{S}$ from the state space $\mathbb{R}^s$ to the action space $\mathbb{R}^a$. The mapping provides a proto-actor $\hat{a}$ in $\mathbb{R}^a$ for a given state under the current parameter. Here, we scale the proto-actor to make sure $\hat{a}$ is a valid action that $\hat{a} \in \mathcal{A}$:
\begin{equation}\nonumber
\mu(s|\theta^{\mu}): \mathcal{S} \to \mathbb{R}^a
\end{equation}
\begin{equation}
\mu(s|\theta^{\mu}) = \hat{a}.
\end{equation}

\emph{K-nearest neighbors:}
The generation of proto-actor can help reduce the computational complexity caused by the large size of the action space. However, reducing the high-dimensional action space to one actor will lead to poor decision making. So, we apply the K-nearest neighbors mapping, $g_{k}$, to expand the actor $\hat{a}$ to a set of valid actions in action space $\mathcal{A}$. The set of actions returned by $g_{k}$ is denoted as $\mathcal{A}_{k}$:
\begin{equation*}
\mathcal{A}_{k}  = g_{k}(\hat{a}_t)
\end{equation*}
\begin{equation}
g_{k} = \mathop{\arg \max}_{a \in \mathcal{A}}^{k} |a-\hat{a}|^2.
\end{equation}

\emph{The critic:}
To avoid the actor with low $Q$-value being occasionally selected, we define a critic network to refine the actor. This deterministic target policy is described below:
\begin{equation}\label{eq:Q}
\hspace{-.3cm}Q(s_t,a_j|\theta^Q) = \mathbb{E}_{r_t,s_{t+1} \scriptsize{\sim} E}[r(s_t,a_t)+\gamma Q(s_{t+1},\mu(s_{t+1}|\theta^{\mu})|\theta^Q)]
\end{equation}
where $\gamma\in\left( 0, 1\right]$ is the discount factor which weigh the future accumulative reward $ Q(s_{t+1},\mu(s_{t+1}|\theta^{\mu})|\theta^Q)$.
Here, the critic takes both the current state $s_t$ and the next state $s_{t+1}$ as its input to calculate the $Q$ value for each action in $\mathcal{A}_k$. Then, the action that provides the maximum $Q$ value will be chosen as $a_t$, i.e., 
\begin{equation}
a_t = \arg \max_{a_j \in \mathcal{A}_{k}} Q(s_t,a_j|\theta^Q)
\end{equation}

\emph{Update:}
The actor policy is updated using deep deterministic policy gradient, which is given as
\begin{equation}
\nabla_{\theta \mu} J \thickapprox \frac{1}{N}\sum_{i}\nabla_a Q(s,a|\mu^Q)|_{s = s_i, a = \mu(s_i)}\nabla_{\theta \mu}\mu(s|\theta^{\mu})|_{s_i}.
\end{equation}
The update of critic network parameter $\theta^{Q}$ and actor network parameter $\theta^{\mu}$ are given as
\begin{equation}
\theta^{Q'} \longleftarrow \tau \theta^{Q} + (1-\tau)\theta^{Q'}
\end{equation}
\begin{equation}
\theta^{\mu'} \longleftarrow \tau \theta^{\mu} + (1-\tau)\theta^{\mu'}
\end{equation}
where $\tau \ll 1$.

\subsection{Workflow}
In this part, we introduce the workflow of the proposed framework. The framework consists of two phases, namely offline and online phases.

\emph{Offline phase:} In the offline phase, the actor and critic networks are constructed and pre-trained with historic transition profiles. This process is the same as in the training of a deep neural network. In the offline phase, when we train the networks with sufficient number of samples, the critic and actor will be sufficiently accurate, and the updated parameters $\theta^{Q}$ and $\theta^{\mu}$ will be stored in order to provide a good initial point for the online phase.

\emph{Online phase:} The online phase is initialized with the parameters determined in the offline phase. The system is dynamically controlled in the online phase. In each epoch $t$, if the requested content is not cached, the DRL agent observes the state $s_t$ from the environment, and obtains the proto actor and $Q$ value from the actor network and critic network, respectively. Then, an $\epsilon$-greedy policy is applied at selecting the execution action $a_t$. This policy can force the agent to explore more possible actions. After the action $a_t$ is executed, the DRL agent observes the reward $r_t$ and next state $s_t$ from the base station cache, and the transition $(s_i, a_i, r_i, s_{i+1})$ will be stored to the memory $\mathcal{M}$ at the end of each epoch. The DRL agent updates the parameters $\theta^{Q}$ and $\theta^{\mu}$ with $N_{\mathcal{B}}$ transition samples from memory $\mathcal{M}$ based on the DDPG.

In our implementation, the actor network has two hidden layers of fully-connected units with 256 and 128 neurons, respectively; and the critic network has two hidden layers of fully-connected units with 64 and 32 neurons, respectively. The capacity of memory $N_{\mathcal{M}}$ is set as $N_{\mathcal{M}} = 10000$, and the mini batch size is set as  $N_{\mathcal{B}} = 100$. The discount factor $\gamma$ introduced in (\ref{eq:Q}) is set as 0.9.

Then, we define the state and action spaces, and the reward function of the DRL agent as follows:

\emph{State Space:}
The DRL agent assumes the feature space of the cached contents and the currently requested content as the state. The feature space consists of three components: short-term feature $\mathcal{F}_s$, medium-term feature $\mathcal{F}_m$, and long-term feature $\mathcal{F}_l$, which represent the total number of requests for each content in a specific short-, medium-, long-term, respectively. These features vary as the cache state is updated. For each decision epoch, we assign a temporary index to every content from which we need to extract features. Since we only extract the features from cached contents and the currently requested content, let the index range from $0$ to the cache capacity $C$. The index of the currently requested content is $0$, while the index of the cached content varies from $1$ to $C$. This temporary index is different from the content ID and is only used for denoting the feature. Then, we let $f_{xj}$, for $x\in \left\lbrace s, m, l\right\rbrace $ and $j\in\left[ 0,C\right] $, denote the feature of a specific content within a specific term. Thus, the observed state is defined as $s_t = \{\mathcal{F}_s; \mathcal{F}_m; \mathcal{F}_l\}$ where
$\mathcal{F}_s=\{f_{s0}, f_{s1},..., f_{sC}\}$,
$\mathcal{F}_m=\{f_{m0}, f_{m1},..., f_{mC}\}$, and
$\mathcal{F}_l=\{f_{l0}, f_{l1},..., f_{lC}\}$.

\emph{Action Space:}
In order to limit the action size, we restrict that the DRL agent can only replace one selected cached content by the currently requested content, or keep the cache state the same. With this, we define $\mathcal{A}$ as the action space, and let $\mathcal{A} = \{0,1,2,..., C\}$, where $C$ is again the cache capacity at the base station. And we assume that only one action can be selected in each decision epoch. Let $a_t$ be the selected action at epoch $t$. Note that, for each caching decision, there are $(C+1)$ possible actions. When $a_t = 0$, the currently requested content is not stored, and the current caching space is not updated. And when $a_t = \upsilon$ with $\upsilon \in \{1, 2,..., C\}$, the action is to store the currently requested content by replacing the $\upsilon^{th}$ content in the cache space.

\emph{Reward:}
As stated in the previous section, we select the cache hit rate as the reward to represent the objective of the proposed framework. The reward for each decision epoch depends on the short and long-term cache hit rate. For example, we set the short-term reward as the number of requests for local content in the next epoch, i.e., the short-term reward $r_t^s$ can be either $0$ or $1$. And let the total number of requests for local content within the next $100$ requests as the long-term reward $r_t^l \in [1,100]$. The total reward for each step is defined as the weighted sum of the short and long-term rewards
\begin{equation*}
r_t = r_t^s + w * r_t^l
\end{equation*}
where $w$ is the weight to balance the short and long-term rewards, so that we can give more priority to the short-term reward to maximize the cache hit rate at every step given the chosen action.

The major notations are listed in Table \ref{tabel:notation} below.
\begin{table}[htbp]
	\caption{Notations}
	\label{tabel:notation}
	\begin{center}
		\begin{tabular}{|l|l|}
			\hline
			Notation & Description                                                                         \\ \hline
			$C$ & Cache capacity at base station \\ \hline
			$i$ & ID, or index of contents \\ \hline
			$N$ & Total number of contents \\ \hline
			$R_t$ & Content requested at epoch $t$ \\ \hline
			$\mathcal{A}$        & Action space     \\ \hline
			$a_t$ & The chosen action in the epoch $t$ \\ \hline
			$r_t$ & The reward obtained in the epoch $t$ \\ \hline
			$s_t$ & The observation state in the epoch $t$ \\ \hline
			$\mathcal{F}$        & Feature space \\ \hline
			$\mathcal{F}_s$, $\mathcal{F}_m$, $\mathcal{F}_l$ & Short/ mid/ long term features \\ \hline
			$f_{si}$, $f_{mi}$, $f_{li}$ & Short/ mid/ long term feature of content $i$ \\ \hline
			$k$        & The number of nearest neighbors              \\ \hline
		\end{tabular}
	\end{center}
\end{table}

	\begin{algorithm}
	\caption{DRL-based Content Caching Algorithm}
	\label{alg:DDPG}
		\begin{algorithmic}[1] 
		\Require
		\State Randomly initialize critic network $Q(s,a|\theta^Q)$ and actor $\mu(s|\theta^{\mu})$ with weights $\theta^{Q}$ and $\theta^{\mu}$.
		\State Initialize target network $Q'$ and $\mu'$ with weights $\theta^{Q'} \longleftarrow \theta^{Q}, \theta^{\mu '} \longleftarrow \theta ^{\mu}$
		\State Initialize replay buffer $\mathcal{M}$ with capacity of $N_{\mathcal{M}}$			
		\State Initialize a random process $\mathcal{N}$ for action exploration
		\State Initialize features space $\mathcal{F}$
		\State Pre-train the actor and critic network with the pairs $<s, a>$ and the corresponding $Q(s,a|\theta^Q)$.
		
		\Ensure
		\For{$t = 1,T$}
		\State The base station receive a request $R_t$
		\If {Requested content is already cached}
		\State Update cache hit rate and end epoch;
		\Else
		\If {Cache storage is not full}
		\State Cache the currently requested content
		\State Update cache state and cache hit rate
		\State End epoch;
		\EndIf
		\State Receive observation state $s_t$
		\State \emph{Actor:}
		\State Receive proto-ation from actor network $\hat{a_t} = \mu(s_t|\theta^{\mu}) $.
		\State \emph{KNN:}
		\State Retrieve $k$ approximately closest actions $\mathcal{A}_{k}  = g_{k}(\hat{a_t})$
		\State \emph{Critic:}
		\State Select action $a_t = \arg \max_{a_j \in \mathcal{A}_{k}} Q(s_t,a_j|\theta^Q) $ according to the current policy.
		\State Execute action $a_t$, and observe reward $r_t$ and observe new state $s_{t+1}$
		\State Store transition $(s_t, a_t, r_t, s_{t+1})$ in $\mathcal{M}$
		\State Sample a random mini batch of $N_{\mathcal{B}}$ transitions $(s_i, a_i, r_i, s_{i+1})$ from $\mathcal{M}$
		\State Set target $y_i = r_i + \gamma Q'(s_{i+1}, \mu '(s_{i+1}|\theta^{\mu '})|\theta^{Q'})$
		\State Update critic by minimizing the loss: $L = \frac{1}{N} \sum_{i}(y_i -Q(s_i,a_i|\theta^Q))^2$
		\State Update the actor policy using the sampled policy gradient:
		\State
		{\small$\nabla_{\theta \mu} J$ $\thickapprox$ $\frac{1}{N}\sum_{i}\nabla_a Q(s,a|\mu^Q)|_{s = s_i, a = \mu(s_i)}\nabla_{\theta \mu}\mu(s|\theta^{\mu})|_{s_i}$}
		
		\State Update the target networks:
		\State \qquad\qquad\qquad
		$\theta^{Q'} \longleftarrow \tau \theta^{Q} + (1-\tau)\theta^{Q'}$
		\State \qquad\qquad\qquad
		$\theta^{\mu'} \longleftarrow \tau \theta^{\mu} + (1-\tau)\theta^{\mu'}$
		\State Update the cache state
		\State Update features space $\mathcal{F}$
		\State Update cache hit rate
		\EndIf				
		\EndFor
		\end{algorithmic}
	\end{algorithm}

\section{Simulation Results}
\subsection{Simulation Setup}
\emph{Data Generation:} In our simulations, the raw data of users' requests is generated according to the Zipf distribution. We set the total number of files as 5000, and we have collected 10000 requests as the testing data. We generate two types of data sets. Initially, we analyze the performance with fixed content popularity distribution, and the data set was generated with unchanged popularity distribution with Zipf parameter set as $1.3$. Subsequently, we study how long-term cache hit rate varies over time as the content popularity distribution changes. In this case, the data set was generated with a varying Zipf parameter, and changing content popularity rank. Note that, although we generate the data using the Zipf distribution, the proposed framework is applicable to arbitrarily distributed popularities, and indeed requires no knowledge regarding the popularity distribution.

\emph{Feature Extraction:} From the raw data of content requests we extract the feature $F$ and use it as the input state of the network. Here, as features, we consider the number of requests for a file within the most recent $10$, $100$, $1000$ requests.

\subsection{Performance Comparison}
To analyze the performance of our algorithm, we evaluate the cache hit rate and provide comparisons with other caching strategies.
\paragraph{Cache Hit Rate}
In this part, comparisons are made between our proposed frame work and the following caching algorithms:
\begin{itemize}
	\item \textbf{Least Recently Used (LRU) \cite{ahmed2013analyzing}:} In this policy, the system keeps track of the most recent requests for every cached content. And when the cache storage is full, the cached content, which is least requested recently, will be replaced by the new content.
	\item \textbf{Least Frequently Used (LFU) \cite{jaleel2010high}:} In this policy, the system keeps track of the number of requests for every cached content. And when the cache storage is full, the cached content, which is requested the least many times, will be replaced by the new content.
	\item \textbf{First In First Out (FIFO) \cite{rossi2011caching}:} In this policy, the system, for each cached content, records the time when the content is cached. And when the cache storage is full, the cached content, which was stored earliest, will be replaced by the new content.
\end{itemize}

Here, we consider both short-term and long-term performance. For the short-term analysis, we study the relationship between the cache capacity and cache hit rate. Regarding the long-term performance, we are interested in the stability and robustness of the proposed DRL framework, i.e., we seek to characterize how the cache hit rate changes over time with the changing popularity distribution of contents.

\begin{figure}
	\centering
	\includegraphics[width=1.1\linewidth]{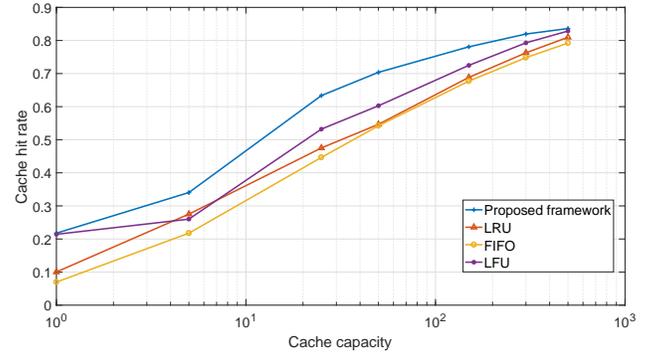}
	\caption{Cache hit rate vs. cache capacity. We vary the cache capacity as $C = 1,5,25,50,150,300,500$.}
	\label{fig:fig1}
\end{figure}

Figure \ref{fig:fig1} shows the overall cache hit rate achieved by the proposed framework and the other caching algorithms introduced above. In this study, we set the Zipf distribution parameter as $1.3$. We can see that our proposed framework provides a higher cache hit rate for all cache capacity values. When the cache capacity is small, the performance of LFU is very close to our proposed framework. As the cache capacity increases, the gap between proposed framework and other three caching algorithms increases at first, and then gradually decreases. At cache capacity $C = 500$, the cache hit rate of all four algorithms are close to each other at around $0.8$. And at this point, the cache hit rates achieved by different policies tend to converge because the cache capacity is high enough to store all popular contents. From this point on, increasing the cache capacity will not improve the cache hit rate effectively any more, and the cache hit rate is now limited by the distribution of the content popularity.

\begin{figure}
	\centering
	\includegraphics[width=1.1\linewidth]{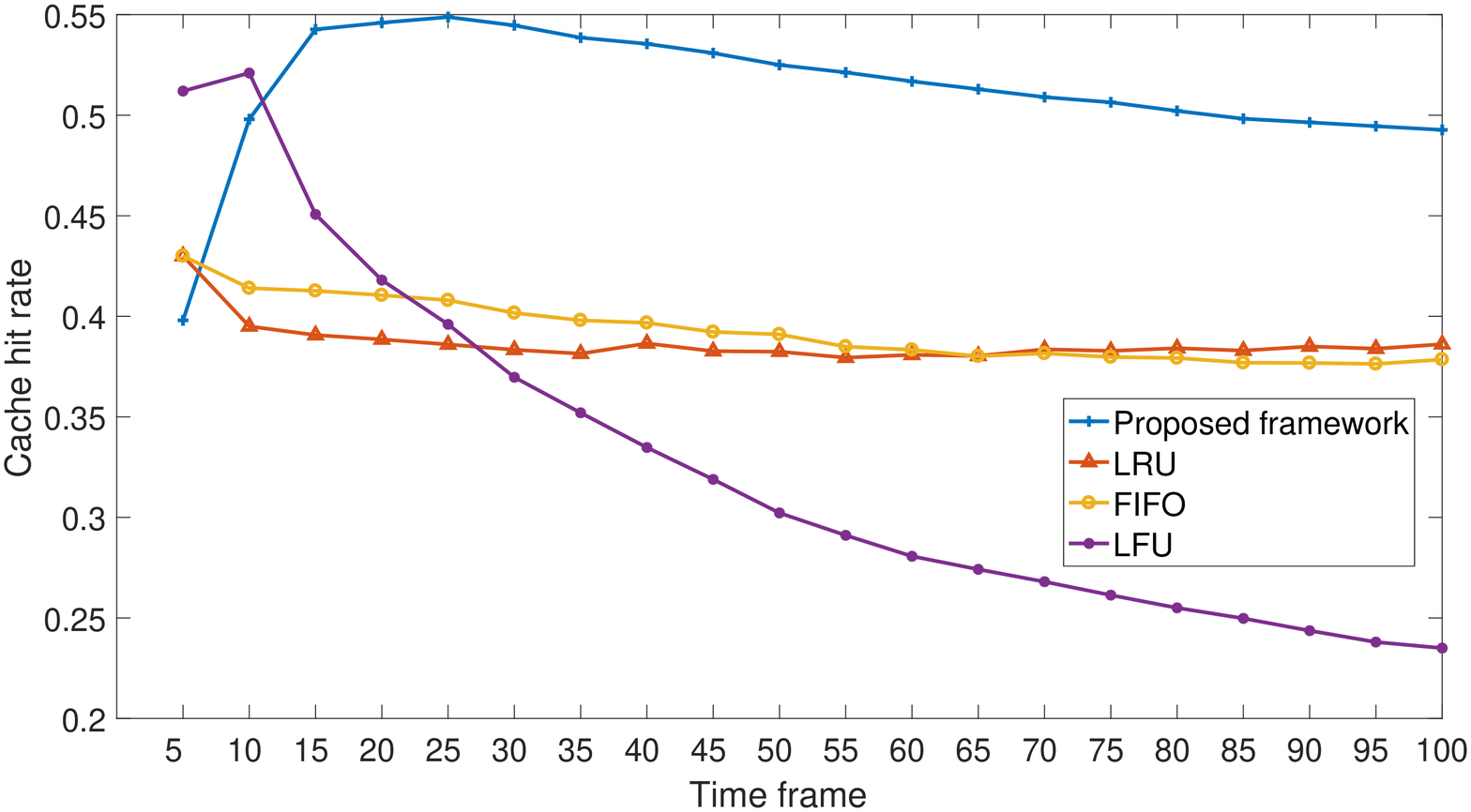}
	\caption{Cache hit rate as the content popularity distribution changes over time, with cache capacity fixed at $C = 300$.}
	\label{fig:fig2}
\end{figure}

In Fig. \ref{fig:fig2}, we address the long-term cache hit rate, and based on the long-term performance we evaluate the capability that the policy can maintain the good performance as content popularities vary over time. Specifically, we design a data set with a changing popularity distribution based on the Zipf distribution. In addition to the parameter of the Zipf distribution, the rank of the contents also vary over time. All the Zipf distribution parameter values and the ranks of contents are generated randomly. From the figure, we can observe that the proposed DRL framework doesn't show advantage initially, but soon the cache hit rate increases. This is because the proposed framework needs to update the deep neural network to adapt to the changing content popularity distribution. After that, the hit rate curve of proposed framework reaches the peak and then deceases only slightly, maintaining a relatively stable cache hit rate. Meanwhile, the LFU curve starts at a relative high cache hit rate and then drops rapidly. This poor performance is caused by the frequency pollution, which is an inevitable drawback of the LFU policy. Because the number of requests are accumulative, when the popularity distribution changes, the previous record will mislead the system. For LRU and FIFO, the performance are relatively stable but the performance is not competitive with respect to our DRL agent. Based on the analysis, our proposed framework will be more suitable for applications that require robustness and a long-term high performance.

\paragraph{Efficiency}
In this part, we compare our proposed framework with the Deep Q-learning based caching algorithm. The most significant difference between these two algorithms is that our proposed algorithm only considers a set of valid actions expanded from the actor, but the Deep Q-learning based algorithm calculates the value for all valid actions. Intuitively, our proposed framework will reduce the computational complexity, but since the Deep Q-learning algorithm receives more possible actions, it may lead to better performance.

\begin{figure}
	\centering
	\includegraphics[width=1.1\linewidth]{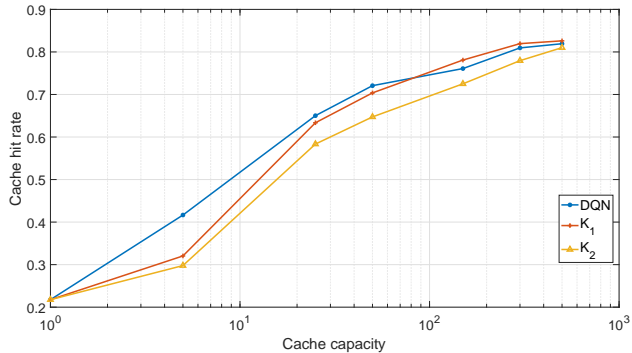}
	\caption{Cache hit rate vs. cache capacity.}
	\label{fig:fig3}
\end{figure}

To address this key tradeoff, we compare the cache hit rates and the corresponding runtimes of these two deep learning schemes. In Fig. \ref{fig:fig3}, the cache capacity values vary as $\{1, 5, 25, 50, 150, 300, 500\}$, and the cache hit rates are plotted when the content requests are generated using the Zipf distribution parameter $1.3$. The curve labeled DQN represents the performance of the deep Q-network. $K_1$ and $K_2$ denote two different settings of proposed framework. In the case of $K_1$, the KNN returns $k_1 = \lceil0.15C\rceil$ actions to the expanded action space $\mathcal{A}_k$. For $K_2$, the KNN returns $k_2 = \lceil0.05C\rceil$ actions to the expanded action space $\mathcal{A}_k$. As we can see in the figure, when cache capacity is $C = 1$, all three curves intersect at the same point, because all three policies are trained to find the one most popular content. Then, as cache capacity increases, the gap between this three policies become obvious. Especially when the cache capacity is $C = 5$, DQN consider all possible actions, while both $K_1$ and $K_2$ only take the proto actor. The gap between $K_1$ and $K_2$ reflects the randomness that might be introduced by the proto action. And then, the gap between $K_1$ and DQN gradually decreases. These results demonstrate that the proposed framework can achieve competitive cache hit rates compared to DQN.

Moreover, the proposed framework can achieve this competitive performance with significantly lower runtimes. With cache capacity fixed at $C = 300$, we record the time needed for $1000$ decision epochs, and show the average runtime results in Table \ref{table:rt} below. As can be seen, the DQN needs much more time at each epoch. In practice, this increased computational cost often leads to storage problems, which makes the deep Q network less competitive in solving large scale problems than the proposed framework.
\begin{table}[htbp]
	\centering
	\caption{Runtime/decision epoch}
	\label{table:rt}
	\begin{tabular}{|l|l|l|l|}
		\hline
		& DQN & $K_1$ & $K_2$ \\ \hline
		Runtime $(s)$ & 1.2225   &  0.3224    & 0.1163    \\ \hline
	\end{tabular}
\end{table}

\section{Conclusion}
In this paper, we have proposed and developed a deep reinforcement learning based content caching policy. We built the framework based on the Wolpertinger architecture and trained it using the deep deterministic policy gradient. We have evaluated the performance of the proposed framework and compared it with both short-term and long-term cache hit rates achieved with LRU, LFU, and FIFO policies. The results show that the proposed framework provides improvements on both short-term and long-term performance. Additionally, we have further confirmed the effectiveness of the proposed framework by comparing the cache hit rate and runtime with those achieved with the deep Q-learning based policy. This comparison has demonstrated that the proposed framework can achieve competitive cache hit rates while effectively reducing the runtime. This makes the proposed framework efficient in handling large-scale data.

This work opens up several directions for future research. First, this current study only considers a scenario involving a single base-station or an access-point. It would be interesting to investigate how to develop a suitable DRL agent for scenarios in which multiple base stations can collaborate with each other. Second, in this work, we assume all contents have the same size, and individual user preferences are not explicitly addressed. It would be practically more appealing to take these factors into consideration. Finally, if both goals can be achieved, we can further develop this framework to address the caching problem in device-to-device communications.

\bibliographystyle{ieeetr}
\bibliography{DRL_caching}

\begin{thebibliography}{10}

\bibitem{kader2015leveraging}
M.~A. Kader, E.~Bastug, M.~Bennis, E.~Zeydan, A.~Karatepe, A.~S. Er, and
  M.~Debbah, ``Leveraging big data analytics for cache-enabled wireless
  networks,'' in {\em Globecom Workshops (GC Wkshps), 2015 IEEE}, pp.~1--6,
  IEEE, 2015.

\bibitem{blasco2014learning}
P.~Blasco and D.~Gunduz, ``Learning-based optimization of cache content in a
  small cell base station,'' in {\em Communications (ICC), 2014 IEEE
  International Conference on}, pp.~1897--1903, IEEE, 2014.

\bibitem{leconte2016placing}
M.~Leconte, G.~Paschos, L.~Gkatzikis, M.~Draief, S.~Vassilaras, and
  S.~Chouvardas, ``Placing dynamic content in caches with small population,''
  in {\em Computer Communications, IEEE INFOCOM 2016-The 35th Annual IEEE
  International Conference on}, pp.~1--9, IEEE, 2016.

\bibitem{li2016streamcache}
W.~Li, S.~M. Oteafy, and H.~S. Hassanein, ``Streamcache: popularity-based
  caching for adaptive streaming over information-centric networks,'' in {\em
  Communications (ICC), 2016 IEEE International Conference on}, pp.~1--6, IEEE,
  2016.

\bibitem{li2016popularity}
S.~Li, J.~Xu, M.~Van Der~Schaar, and W.~Li, ``Popularity-driven content
  caching,'' in {\em Computer Communications, IEEE INFOCOM 2016-The 35th Annual
  IEEE International Conference on}, pp.~1--9, IEEE, 2016.

\bibitem{song2017learning}
J.~Song, M.~Sheng, T.~Q. Quek, C.~Xu, and X.~Wang, ``Learning based content
  caching and sharing for wireless networks,'' {\em IEEE Transactions on
  Communications}, 2017.

\bibitem{tanzil2017adaptive}
S.~S. Tanzil, W.~Hoiles, and V.~Krishnamurthy, ``Adaptive scheme for caching
  youtube content in a cellular network: Machine learning approach,'' {\em IEEE
  Access}, vol.~5, pp.~5870--5881, 2017.

\bibitem{gruslys2017reactor}
A.~Gruslys, M.~G. Azar, M.~G. Bellemare, and R.~Munos, ``The reactor: A
  sample-efficient actor-critic architecture,'' {\em arXiv preprint
  arXiv:1704.04651}, 2017.

\bibitem{lillicrap2015continuous}
T.~P. Lillicrap, J.~J. Hunt, A.~Pritzel, N.~Heess, T.~Erez, Y.~Tassa,
  D.~Silver, and D.~Wierstra, ``Continuous control with deep reinforcement
  learning,'' {\em arXiv preprint arXiv:1509.02971}, 2015.

\bibitem{dulac2015deep}
G.~Dulac-Arnold, R.~Evans, H.~van Hasselt, P.~Sunehag, T.~Lillicrap, J.~Hunt,
  T.~Mann, T.~Weber, T.~Degris, and B.~Coppin, ``Deep reinforcement learning in
  large discrete action spaces,'' {\em arXiv preprint arXiv:1512.07679}, 2015.

\bibitem{ahmed2013analyzing}
M.~Ahmed, S.~Traverso, P.~Giaccone, E.~Leonardi, and S.~Niccolini, ``Analyzing
  the performance of {LRU} caches under non-stationary traffic patterns,'' {\em
  arXiv preprint arXiv:1301.4909}, 2013.

\bibitem{jaleel2010high}
A.~Jaleel, K.~B. Theobald, S.~C. Steely~Jr, and J.~Emer, ``High performance
  cache replacement using re-reference interval prediction (rrip),'' in {\em
  ACM SIGARCH Computer Architecture News}, vol.~38, pp.~60--71, ACM, 2010.

\bibitem{rossi2011caching}
D.~Rossi and G.~Rossini, ``Caching performance of content centric networks
  under multi-path routing (and more),'' {\em Relat{\'o}rio t{\'e}cnico,
  Telecom ParisTech}, pp.~1--6, 2011.

\end{thebibliography}

\end{document}